\documentclass[useAMS,usenatbib]{mn2e}
\usepackage{amsmath}    
\usepackage{amssymb}
\usepackage{journals}
\usepackage{graphicx,dblfloatfix}   
\usepackage{verbatim}   
\usepackage{color}      
\usepackage{ulem}		
\usepackage{slashbox}
\usepackage[section]{placeins}
\newcommand{\sgra}{Sgr~A$^{*}$}

\title[Using infrared/X-ray flare statistics to probe the emission regions near the ISCO of \sgra]{Using infrared/X-ray flare statistics to probe the emission regions near the event horizon of \sgra}
\author[Salom\'e Dibi et al.]{S. Dibi$^{1}$\thanks{e-mail:
s.dibi@astro.ru.nl},  S. Markoff$^{2}$, R. Belmont$^{3,4}$,  J. Malzac$^{3,4}$, J. Neilsen$^5$, G. Witzel$^6$\\\\
$^{1}$Department of Astrophysics/IMAPP, Radboud University, P.O. Box 9010, 6500 GL
Nijmegen, The Netherlands\\
$^{2}$Anton Pannekoek Institute of Astronomy, University of Amsterdam, Postbus 94249,
1090 GE Amsterdam, The Netherlands\\
$^{3}$Universit\'e de Toulouse; UPS-OMP; IRAP; Toulouse, France \\
$^{4}$CNRS; IRAP; 9 Av. colonel Roche, BP 44346, F-31028 Toulouse cedex 4, France \\
$^{5}$MIT Kavli Institute for Astrophysics and Space Research, 77 Massachusetts Ave, NE80-6079 Cambridge, MA 02139, USA\\
$^{6}$Dep. of Physics and Astronomy, University of California Los Angeles (UCLA), 465 Portola Plaza, Los Angeles, CA 90095, USA}

\begin{document}

\date{Accepted in MNRAS on June 8, 2016} 

\pagerange{\pageref{firstpage}--\pageref{lastpage}} \pubyear{2016}

\maketitle

\label{firstpage}

\begin{abstract} 
The supermassive black hole at the centre of the Galaxy flares at least daily in the infrared (IR) and X-ray bands, yet the process driving these flares is still unknown. So far detailed analysis has only been performed on a few bright flares. In particular, the broadband spectral modelling suffers from a strong lack of simultaneous data. However, new monitoring campaigns now provide data on thousands of flaring events, allowing a statistical analysis of the flare properties. In this paper, we investigate the X-ray and IR flux distributions of the flare events.

Using a self-consistent calculation of the particle distribution, we model the statistical properties of the flares. Based on a previous work on single flares, we consider two families of models: pure synchrotron models and synchrotron self-Compton (SSC) models. We investigate the effect of fluctuations in some relevant parameters (e.g. acceleration properties, density, magnetic field) on the flux distributions. The distribution of these parameters is readily derived from the flux distributions observed at different wavelengths.

In both scenarios, we find that fluctuations of the power injected in accelerated particles plays a major role. This must be distributed as a power-law (with different indices in each model). In the synchrotron dominated scenario, we derive the most extreme values of the acceleration power required to reproduce the brightest flares. In that model, the distribution of the acceleration slope fluctuations is constrained and in the SSC scenario we constrain the distributions of the correlated magnetic field and flow density variations.

\end{abstract}


\section{Introduction}
Sagittarius A* provides our best opportunity to study black hole accretion because of its size and proximity. This supermassive black hole is a faint source emitting at several wavelengths. It was first observed in the radio wavelengths by \citet{balick74} and named Sagittarius A* (\sgra) by Robert Brown in 1982 because it is the brightest radio source in the Sagittarius region of the sky. 

The flaring behaviour observed from this source is crucial for our understanding of the physics at the event horizon scale. It provides detailed information about the microphysics driving the plasma close to the black hole. An accurate model for the microphysics is even more important as our observations are reaching a very high level of resolution (especially with the Event Horizon Telescope returning results, e.g. \citealt{fish16}). Interpreting the high quality data requires taking into account particle distributions in the medium and understanding what the physical plasma conditions are. Studying flares allows us to answer these questions by describing the distribution of key plasma parameters.

Intensive observational and theoretical studies have been performed over the last decades, leading to many solid constraints for the physical properties of \sgra's. The mass of our central black hole has been estimated to an accuracy of a few percent by tracking the orbits of the stars around it \citep{schodel02, eisenhauer05, melia07, ghez08, gillessen09}. \sgra 's observed luminosity is very low, with a bolometric value of only few hundred solar luminosities; for a black hole mass of $ 4.3\times 10^6 \ M_{\rm \odot}$, this corresponds to $\sim  10^{-9} \ L_{\rm Edd}$. This weak emission can be explained in terms of a radiatively inefficient accretion flow (\citealt{yuan14}, and references therein) and can also be fitted by jet models \citep{Falcke93, falcke-markoff00, yuan02, moscibrodzka13}. 
\sgra's accretion rate is very low ( $ \dot{M}<2\times10^{-7} \ M_{\rm \odot}~{\rm yr^{-1}}$) as deduced from observations of polarization and Faraday rotation measurements \citep{bower03,marrone07}. This accretion rate is consistent with general relativistic magnetohydrodynamical (GRMHD) simulations that estimate a rate of a  few times $10^{-9} \ M_{\rm \odot}~{\rm yr^{-1}}$ \citep{moscibrodzka09, drappeau13}. The bulk of the emission from \sgra\ is observed at radio and submillimeter frequencies, reaching a maximum of about 1 Jansky  at the peak of the spectrum (e.g. \citealt{falcke98, melia01}). The optical and UV bands are completely obscured by interstellar extinction, but it is detected in the infrared (IR) and X-ray bands. The IR emission from the source itself is difficult to observe because of the contamination by many stars in the central region. Nevertheless, adaptive optics  led to a conclusive IR identification of Sgr A* (e.g. \citealt{genzel03, schodel11, shahzamanian15}). Even though \sgra\ is the least luminous black hole ever observed (in Eddington-scaled units), radiative losses can still play a role in the dynamics of the system above an accretion rate of $10^{-8} \ M_{\rm \odot}~{\rm yr^{-1}}$ \citep{dibi12}. The extremely low luminosity of \sgra\ is also an opportunity to observe intense and intriguing variability that is not observed in such a way in any other black hole system on comparable timescales.

The flux variability in the radio and submillimeter bands is on the order of 20$\%$ or less \citep{marrone08,yusefzadeh10}. However, in the IR and X-ray wavelengths, the spectrum is highly variable on time scales of a few thousand seconds. 

The NIR emission is linearly polarized, strongly suggesting that the synchrotron process is responsible for the IR emission \citep{eckart06,shahzamanian15}. Observations and models agree on the fact that synchrotron emission by non-thermal relativistic electrons can account for the IR flares, but what exactly triggers electron acceleration is not modelled in detail and is therefore subject to interpretation (see e.g.   \citealt{markoff01,yuan03,tagger06,yuan09,doddseden10,zubovas12,dibi14}).

In the X-ray band, flares are observed to occur about once a day on average, with luminosities reaching a few times $1\times 10^{35}\rm  erg \ s^{-1}$ for the brightest events (more than two orders of magnitude above the quiescent background level) with an average flare timescale of an hour or less. Because $\sim$90\% of this background originates at large radii (extending to the Bondi radius; \citealt{neilsen13, neilsen15}), the brightest flares may be over 1000 times more luminous than the quiescent inner accretion flow. The quiescent X-ray emission has been observed by the \textit{Chandra X-ray Observatory}, and many flares have been observed by Chandra, XMM Newton, Swift, and NuSTAR \citep{baganoff01,baganoff03,nowak12,degenaar13, neilsen13, barriere14}. Recently, the X-ray satellite NuSTAR has extended the window for flare observations into the hard X-rays. \citet{barriere14} reported a variation of the photon index between two distinct flares of similar flux from $\alpha=2.84^{+0.64}_{-0.54}$ \ to \ $\alpha = 2.04^{+0.22}_{-0.20}$ \ at $95\%$ confidence. This result shows that whatever mechanism is responsible for the particle acceleration in the flares does not act with the same efficiency each time.

Due to the scarcity of purely simultaneous observations, the correlations between IR and X-ray flares are still poorly constrained. At least a few flares have been identified in both bands and clearly associated (within a few minutes). These observations tend to show that the strongest NIR flares are generally associated with an X-ray flare \citep{schodel11,bremer11,doddseden11,genzel03, ghez04} and that they last longer \citep{doddseden09, yusefzadeh12, eckart12}. Unfortunately, broad observational coverage of these simultaneous flares is very rare and has not allowed a clear understanding of the flare physics yet.

So far, most of the flare properties have been derived from the modelling of a few non-simultaneous bright flares (i.e. no multiwavelength coverage of the event). Two kinds of models have been used to model their broadband emission: a) {\it Synchrotron-dominated models (SD)}, where the emission is entirely due to synchrotron radiation of accelerated particles, from IR to X-rays. The spectrum of such a model model is a broken power-law with a cooling break. Below the break, acceleration competes with particle escape, while at higher energy, it competes with synchrotron cooling. b) {\it Synchrotron Self-Compton models (SSC)}, where the IR emission is due to synchrotron radiation, while the X-ray emission results from Comptonisation of the synchrotron photons by the same electrons. Although synchrotron-dominated models were favoured in the modelling of individual flares \citep{dibi14}, SSC models could not be ruled out.

With the increasing number of observations and new monitoring campings, we are now able to quantify and model the statistical properties of the flare distributions. The IR flux distribution has been studied and reported by \citet{doddseden10} and \citet{witzel12} using data from NACO/VLT in the Ks band ($\lambda = 2.18 \mu m$, $\nu=1.1\times 10^{14}$ Hz). It was built by measuring the flux in 1 min long intervals. In both works the number of events is found to decrease with flux. However, there is no clear agreement on the precise distribution shape. Analysing data between 2004 and 2009, \citet{doddseden10} used a two-component distribution to model the data, with a log-normal population of low flux flares and a power-law distribution with index  $\Gamma=-2.7\pm 0.14$ at higher fluxes. In contrast, \citet{witzel12}, analysed data between 2003 and 2010 with a different reduction method and found a single power-law distribution with index \ $\Gamma=-4.2\pm 0.1$ at all fluxes\footnote{We note that the mentioned indices correspond to the flux distributions and not the cumulative flux distributions.}. More recently, \citet{meyer14} also argued against the existence of two distinct states. In the present paper we assume that the flux distribution follows the statistics found by \citet{witzel12} and supported by \citet{meyer14}.

Thanks to the \textit{Chandra} 2012 X-ray Visionary Project\footnote{http://www.sgra-star.com/} dedicated to \sgra, we have tripled the number of flare events observed in the 2-8 keV band and now have enough data to derive their flux distribution. 
This has been done in two ways. \citet{neilsen13} considered only well resolved individual flares and assumed a common power-law spectral shape. They find that the averaged flare luminosity is distributed as a power-law with index $\Gamma=-1.9^{+0.3}_{-0.4}$. This result was then confirmed by \citet{neilsen15} who built the flare distribution in 300s long time intervals (without any flare shape assumptions) and found a power-law with index $1.92_{-0.02}^{+0.03}$. 

In this paper, we aim at interpreting these new observational results and modelling both the IR and X-ray flux distribution using a single, broad-band model.
Although the observed flux distributions are derived from thousands of data points, there are still very few flares that are well-covered at both wavelengths. This presents a challenge for broadband analysis. However, a statistical study can provide results to complement the study of individual flares. While the statistical analysis of the flux distributions does not provide any constraint on the timing properties of the flares (discarding precise information about flare evolution, variability, and frequency-dependent delays), it enables the modelling of statistical properties of the numerous weak flares: the models must not only be able to reproduce a few isolate flares, but also the large range of fluxes observed in different bands.  As we will show, combining the flux distribution in IR and  X-rays provides new constraints on the statistical properties of the physical parameters responsible for the flux variability. 


The paper is organised as follows. The flare model and the method of building flux distributions are described in in section \ref{Method_distribution}. The results of the statistical analysis within the framework of two different scenarios (SD and SSC) are presented in section \ref{Results1} and \ref{Results2} respectively.

\section{Model}
\label{Method_distribution}

The method to model the cumulative distributions of fluxes (CDFs) has two steps. First we use an emission model to produce the spectrum resulting from a given parameter set. Then, we vary the parameters according to given probabilities and compare them to the observed CDFs. 

\subsection{Emission model}
The short variability time scales indicate that the emission originates from a very small region in the innermost parts of the accretion flow. Here we assume that it is produced by a lepton population in a uniform and isotropic region of size $R=2 \ r_{\rm G} = 1.3 \times 10^{12}$ cm for \sgra, where $\ r_{\rm G}=GM/c^2$ is the gravitational radius of the black hole which has a light crossing time of about 21 s. The emission is modelled with the {\sc belm} code \citep{belmont08} that simultaneously solves the coupled kinetic equations for leptons and photons in a magnetized plasma. The microphysics implemented includes radiative processes such as self-absorbed synchrotron, Compton scattering, self-absorbed bremsstrahlung radiation, pair production/annihilation, Coulomb collisions, and prescriptions for particle heating/acceleration. The emission is essentially governed by the following physics: 
\begin{itemize}
\item The magnetic field $B$ is described by the dimensionless compactness parameter: $l_b= (\sigma_T RB^2)/(8\pi m_e c^2)$. 
\item As matter accretes inwards, particles can be injected into and then escape from the emission region. The incoming plasma is assumed to enter the emission region with temperature: $\theta_{\rm inj}=k_BT_{inj}/(m_ec^2)= 13$ and rate $\dot{n}_{\rm inj}$ (electron cm$^{-3}$s$^{-1}$). The injection is parameterized with compactness $l_{\rm inj} \approx 3 \theta_{\rm inj} R^2 \sigma _T \dot{n}_{\rm inj}/c$. Particles are assumed to escape on a typical time scale $t_{\rm esc}=R/c$.
\item Particles can also be accelerated locally by some non-thermal acceleration process (e.g., shocks and/or reconnection). In the {\sc belm} code, the acceleration mechanism is mimicked by taking particles uniformly from the distribution and re-injecting them at high energy as a power-law, thus conserving the particle number in the modelled region. It is parametrised by the injected power $L_{nth}$ (described by the corresponding compactness parameter $l_{\rm nth}= (\sigma_TL_{\rm nth})/(Rm_ec^3)$) and the properties of the power-law. Following \citet{dibi14} we use a minimum Lorentz factor $\gamma_{\rm min}=50$ corresponding to the typical energy of the thermal population, and a maximal energy of $\gamma_{\rm max}=4.6\times 10^5$, large enough to reproduce the high energy X-ray flares observed by NuSTAR \citep{barriere14}. The power-law index $s$ is expected to be around $s\approx2$ for diffusive shock acceleration \citep{bell78, rieger07, guo14} but is a free parameter of the model. 
\end{itemize}
When written in terms of these dimensionless parameters, the equations and results are only weakly sensitive to the exact value of the region size $R$ (only the synchrotron self-absorption frequency depends explicitly on this length scale). In the above description,  $\sigma_T$ is the Thomson cross section, $k_B$ is the Boltzmann constant, $m_e$ is the electron mass, and $c$ is the speed of light. There are only 4 free parameters to the model: $l_b$, $l_{\rm nth}$, $s$, and $l_{\rm inj}$, while the other parameters are set to standard values based on decades of multiwavelength fits from the quiescent spectrum (see \citealt{dibi14} and references therein). The prescribed acceleration process competes with injection, escape, and cooling losses to give the resulting particle distribution from which the emission is calculated. 

The {\sc belm} code was recently used to model the emission of two bright X-ray flares \citep[2012 July 21th, and 2012 October 17th, ][]{dibi14} with additional constraints from (non-simultaneous) IR flares. Two classes of models were studied: SD and SSC models. It was shown that it is possible to reproduce the quiescent state and the observed flares in the framework of both models. The value of the free parameters are summarised in Table \ref{tab:model_parameters} for each of these models. In the SD  scenario, changes in $s$ and $l_{\rm nth}$ were sufficient to reproduce both the quiescent and the flaring state. In contrast, from the same quiescent state, changes in all four free  parameters were required to reproduce the flares in the SSC case. Here we produce synthetic CDFs in the framework of these two models and compare them to the observations. 

\begin{table}
 \begin{center}
 \begin{tabular}{lcccc}
  \hline
 & $B$  &  $l_{\rm nth}$ &  $l_{\rm inj}$  &  $s$   \\ 
   &(G)  &  ($\times 10^{-5}$) &   ($\times 10^{-4}$) &   \\ 
\hline
Quiescence & 175.3  &  1 & 4.64 &  3.60 \\
Flare: PS & 175.3 &   9.8/11.4 & 4.64 & 2.28/2.13 \\
Flare: SSC & 34.5 & 100 & 200 &  2.60\\
\hline
\end{tabular}
  \caption{Parameter values in the quiescent state and in the flaring state for the two models considered in \citet{dibi14}: SD and SSC. The columns give the magnetic field strength, the non-thermal and injection compactness parameters and the slope of the accelerated particle distribution. The two different values given for $l_{nth}$ and $s$ in the Flare:PS case refer to the reproduction of two distinct very bright flares observed by NuSTAR.} \label{tab:model_parameters}
\end{center}
\end{table}


\subsection{Flux distributions}
\label{Generation of flare distributions}


Here we suppose that the particle distributions and emitted spectra are always in a quasi-steady state, and that the observed variability in both bands is a direct result of variations in one (or more) of these parameters. This assumption is supported by the fact that the cooling time scales are typically very short (see \citealt{dibi14}). In this paper we investigate several probabilistic models for the distributions of these parameters.

To produce synthetic CDFs, we draw a large number of random values for the varying parameters according to given probabilities. If a parameter has probability $p(x)$ to have a value $x$ in range $[x_{\rm min},x_{\rm max}]$, its distribution can be produced numerically by drawing random numbers $\xi$ with uniform distribution between 0 and 1 and each time computing the value $x$ such that $\int _{x_{\rm min}} ^{x} p(x')dx' = \xi$. We draw typically 10 000 random values for each varying parameter.

For each set of parameters, we compute the corresponding spectrum using the {\sc belm} code, derive the IR (at 2.8 $\mu$m) and X-ray (between 2 and 8 keV) fluxes, and build the associated flux distributions. Running the code for the thousands of random parameter sets is time consuming. Instead, we first build tables of spectra (with typically 100 values per parameter) and then, for each random set of parameters, extract the spectra corresponding to the closest parameter values in those grids. 

For direct comparison to observational results, we compute the CDF of the flux, i.e. the fraction of all fluxes that are greater than or equal to a given flux.

\section{Synchrotron-dominated scenario}
\label{Results1}

In this section we explore the scenario where the flare emission is the result of non-thermal synchrotron emission with a cooling break. This scenario is good at reproducing the (non-simultaneous) multiwavelength data, has the simplest assumptions, and the plasma parameters are in good agreement with the current knowledge of \sgra 's environment in terms of density, magnetic field value, and temperature. Moreover, only the particle acceleration properties need to be varied to reproduce the quiescent and flaring states \citep{dibi14}.  
Within this synchrotron scenario, a large table of spectra is generated by varying the non-thermal compactness and the acceleration slope according to table \ref{tab:PS_grid} while the other parameters are set to $l_{\rm inj}=4.6\times10^{-4}$ (leading to a density of about $3.3\times 10^6$  cm$^3$) and $B=175$ G. 
\begin{table}
\begin{center}
\begin{tabular}{c|cccc}
&min &max &number& grid\\ \hline
$l_{\rm nth}$ & $5\times 10^{-6}$ & $5 \times 10^{-3}$ & 1000 & log \\
 $s$ & 1 & 4  & 100 & lin \\
 \hline
\end{tabular}
\caption{Properties of the grids used to produce the table of spectra in the synchrotron dominated model.} \label{tab:PS_grid}
\end{center}
\end{table}
This range of values has proven to be capable of reproducing the multiwavelength spectra from the quiescent to the flaring state.

\subsection{IR and X-ray fluxes}
\begin{figure}
\includegraphics[scale=0.35]{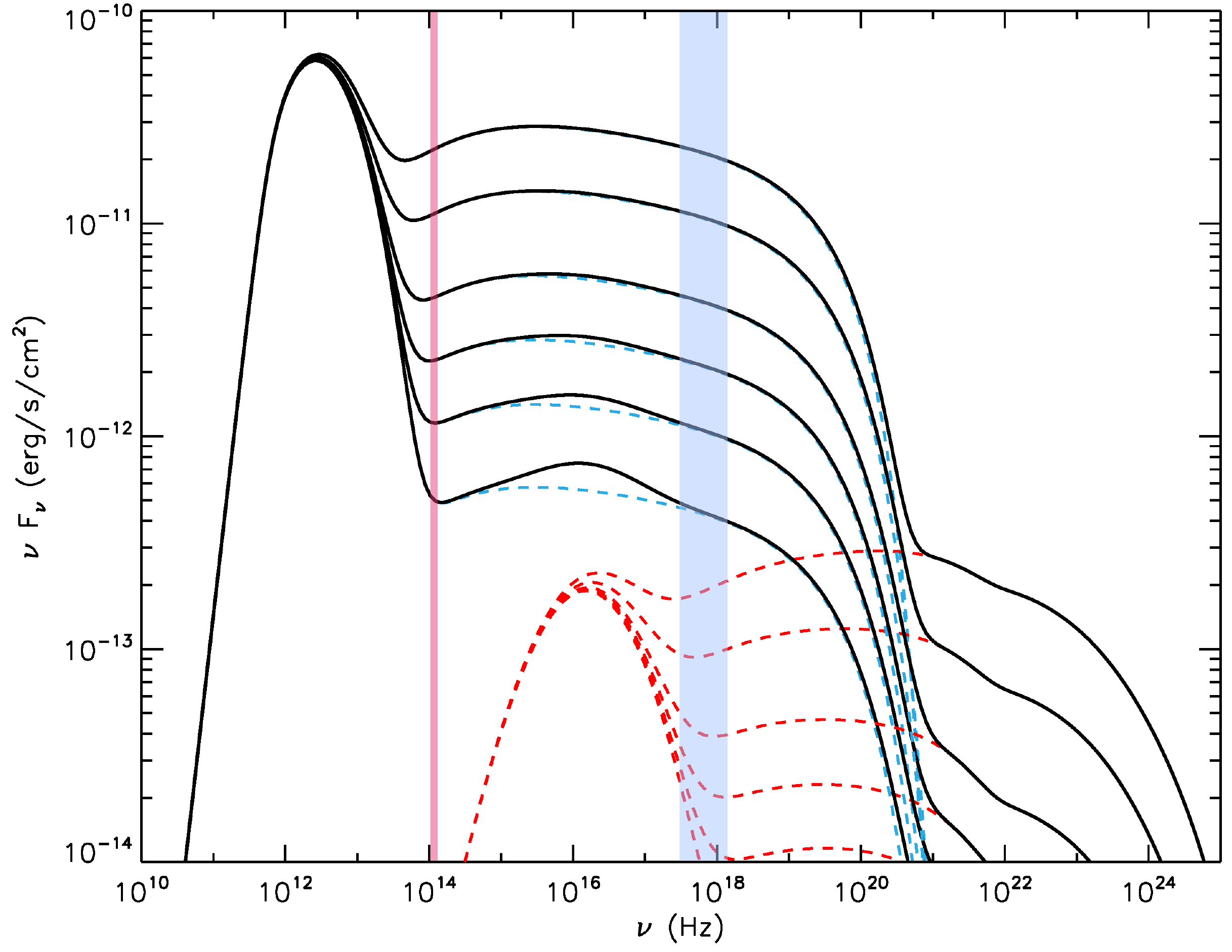}
\includegraphics[scale=0.35]{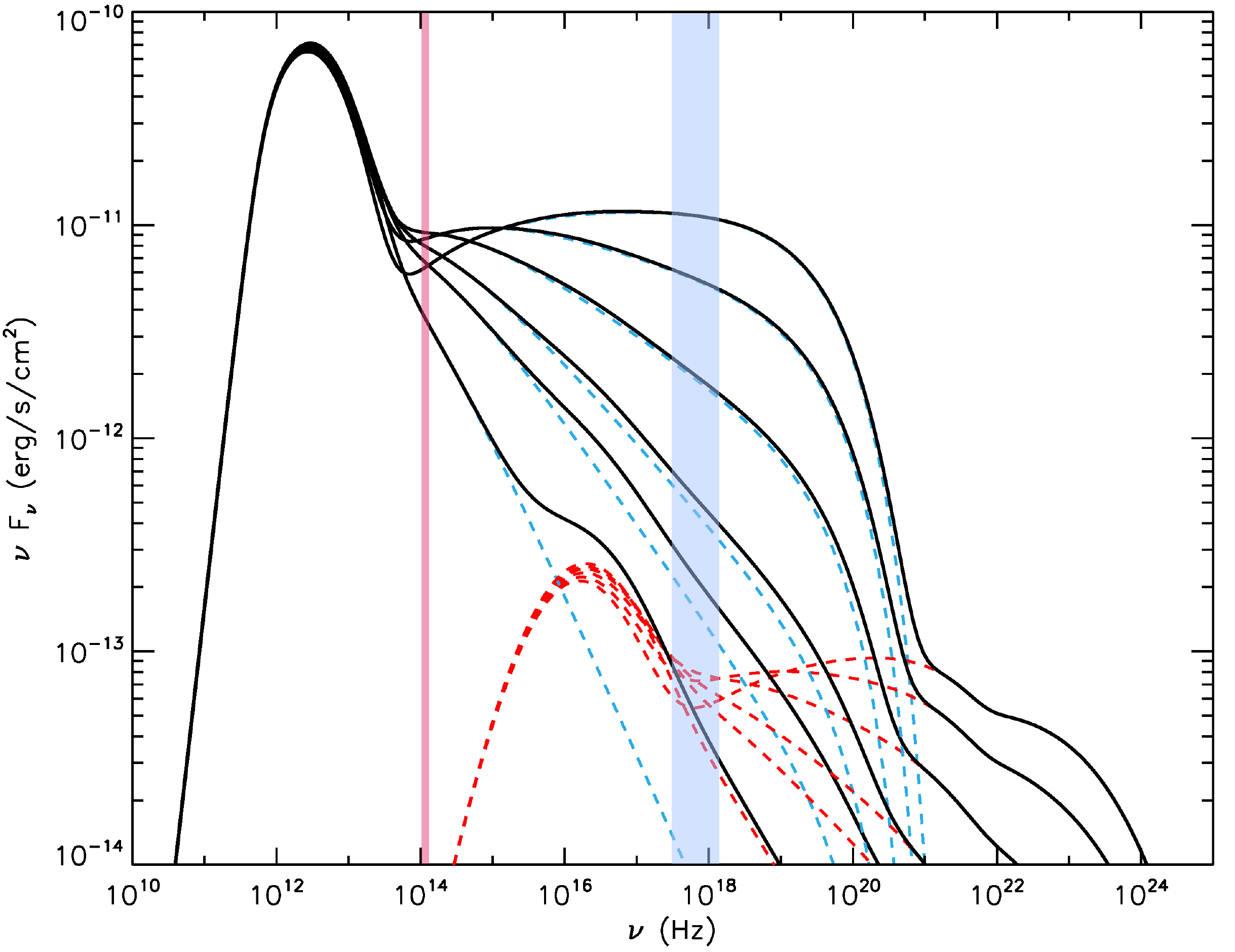}
\caption{Effect of the varying parameters on synchrotron-dominated spectra. {\bf Upper panel:} Effect of non-thermal power: $l_{nth}=2 \times 10^{-6}, 5\times 10^{-6}, 1 \times 10^{-5}, 2\times 10^{-5}, 5 \times 10^{-5},  1 \times 10^{-4}$ from bottom to top (with $s=2.13$). {\bf Lower panel:} Effect of the acceleration slope $s=3.6, 3.0, 2.8, 2.5, 2.2, 1.8$ from bottom to top (with $l_{nth}= 3.2 \times 10^{-5}$). The black line is the total spectrum while the underlying dashed blue and red lines show the synchrotron and Compton contributions.The vertical light red and blue bands indicate for the Ks and 2-8keV bands respectively.}
\label{Spectra_lnth}
\end{figure}

The effect of the two parameters of this model is illustrated in Fig. \ref{Spectra_lnth}. As long as the acceleration process is not too efficient and only a small fraction of the thermal pool is accelerated to high energy, the synchrotron emission simply scales with the acceleration power in both bands, i.e., $F_X\propto F_{IR} \propto l_{\rm nth}$. Because the spectrum pivots around the IR band when the acceleration slope s is varied, the IR flux is relatively insensitive to the value of $s$ and is thus simply proportional to $l_{\rm nth}$:
$$ F_{IR} \propto l_{\rm nth} \ .$$ 

In contrast, the X-ray flux is very sensitive to the acceleration flux. Once accelerated with slope $s$, the electrons cool through synchrotron radiation and reach a steady state distributions with slope $s+1$, which corresponds to a power-law synchrotron spectrum with flux index $\alpha=(s+1)/2$.
Hence, the synchrotron flux at a given band is proportional to $A^{-s}$, where, at a given frequency, $A$ is a constant.
Due the strong lever arm effect, a small change in $s$ produces a large change in the X-ray flux. As long as the acceleration slope is not too soft, the Compton contribution remains negligible in the X-ray band and the total flux is synchrotron dominated. Then, the X-ray flux depends on the varying parameters as:
$$ F_{X} \propto l_{\rm nth}A^{-s} \ .$$ 

\subsection{IR CDF}
\begin{figure}
\begin{center}
\includegraphics[width=\columnwidth]{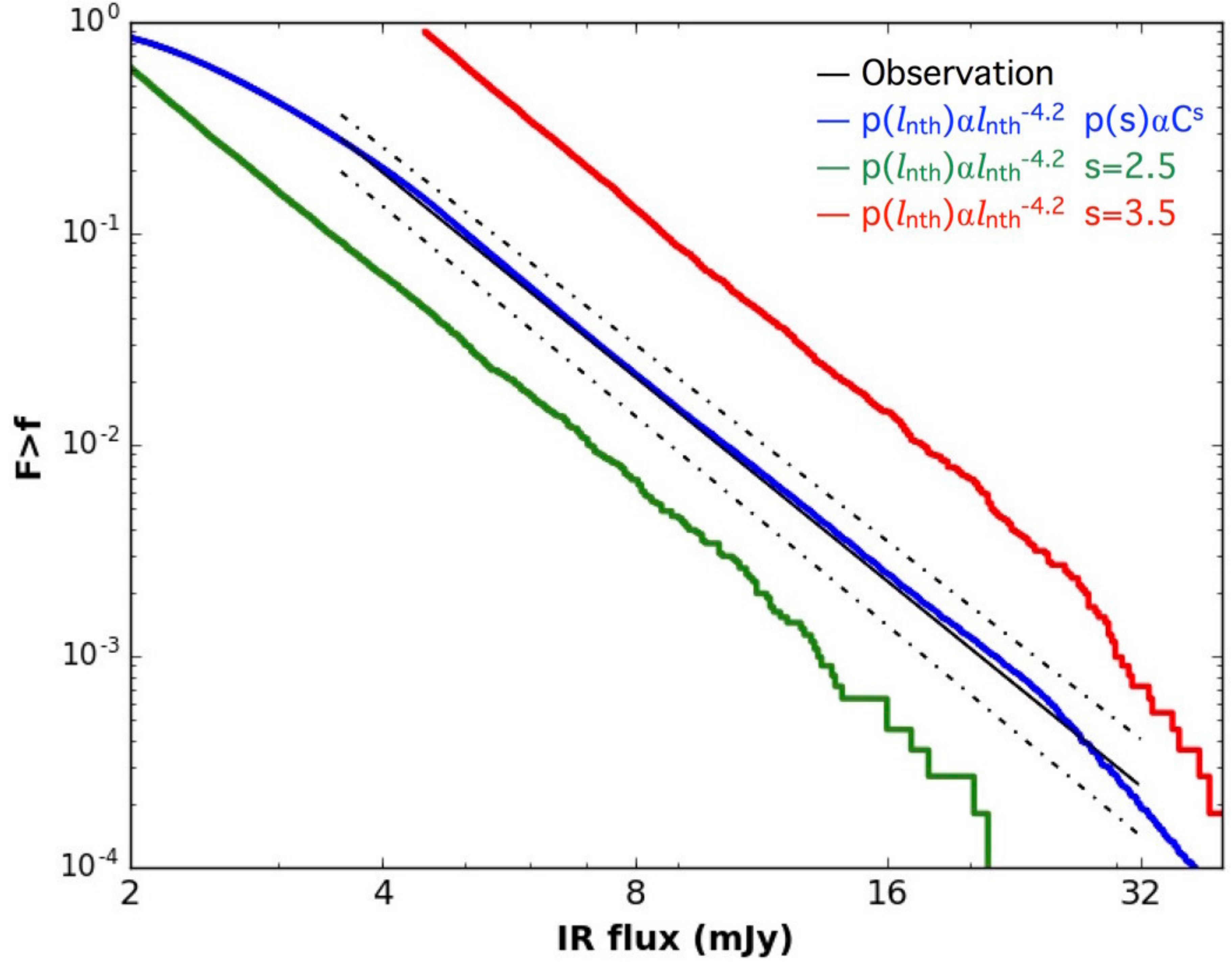}
\caption{The three simulated IR CDF are produced by the non-thermal probability $p(l_{\rm nth}) \propto l_{\rm nth}^{-4.2}$ in the synchrotron-dominated scenario. The red and green curve have constant acceleration slope of $s=2.5$ and $3.5$, respectivelly. The blue curve has a probability distribution for the slope of $p(s)\propto C^{s}$.} \label{fig:PS:IR_CDF}
\end{center}
\end{figure}

As the IR flux is only sensitive to the acceleration power, the observed IR flux distribution gives a direct constraint on the statistical distribution of this parameter. As the IR flux scales linearly with $l_{\rm nth}$, the distribution of non-thermal compactness that produces a power-law flux distribution with index $\Gamma=-4.2$ is a power-law with same index: 
$$p(l_{\rm nth}) \propto l_{\rm nth}^{-4.2} \ .$$ 
The synthetic IR CDF produced by this distribution is shown in Fig. \ref{fig:PS:IR_CDF} for different values of the acceleration slope.
As expected, the found CDF is insensitive to the acceleration slope.

The CDF observed in IR does not show any evidence for a cutoff at large flux so the extremal values of the parameters cannot be directly inferred. However, the observed distribution extends at least to $F_{IR} = 30$ mJy \citep{witzel12}, corresponding to compactness values as large as $l_{\rm nth}^{\rm max} > 1\times 10^{-4}$ and to an acceleration power of: $L_{\rm nth}^{\rm max}>4.8\times 10^{36}$ erg/s.

\subsection{X-ray CDF}
The X-ray flux depends both on the non-thermal power and the acceleration slope. If $s$ is kept constant, the X-ray flux distribution is the same as the IR distribution (i.e. a power-law with index 4.2). Such a steep distribution is incompatible with the observed flux distribution with index $\Gamma=1.9$ \citep{neilsen15}. 

\begin{figure}
\begin{center}
\includegraphics[width=\columnwidth]{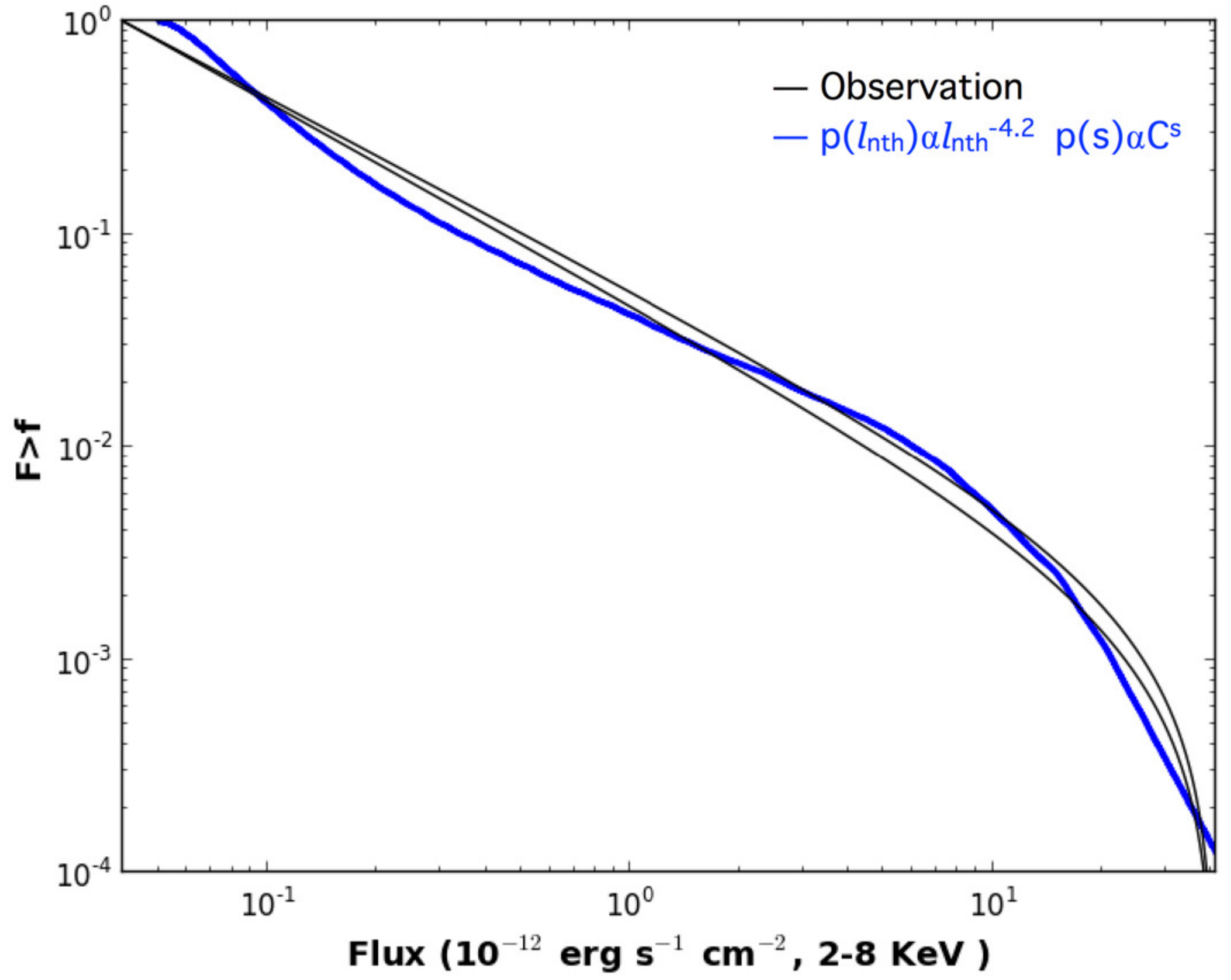}
\caption{The blue line is the X-ray CDF produced  in the synchrotron-dominated scenario by independent parameters $l_{nth}$ and $s$ with probability distributions $p(l_{\rm nth}) \propto l_{\rm nth}^{-4.2}$ and $p(s)\propto C^{s}$ respectively. The black lines represent the inferred power law distribution of the variable portion of the X-ray emission (\citealt{neilsen15}).} \label{fig:PS:X_CDF}
\end{center}
\end{figure}

As the $l_{\rm nth}$ distribution is well constrained by the IR CDF, it is in principle possible to constrain the precise $s$ distribution from the X-ray CDF. This requires additional assumptions about the $s-l_{\rm nth}$ correlation. For instance, if the two parameters are independent, we find empirically that the X-ray CDF is well reproduced with the probability: $p(s) \propto C^{s}$ with $C=66$, as shown in in Fig. \ref{fig:PS:X_CDF}. However, it is not clear whether these two acceleration properties are independent. Nonetheless, regardless of the $s-l_{\rm nth}$ correlation, it is interesting to note that X-ray flares are observed to at least $F_{X}>10^{-11}$ erg s$^{-1}$ cm$^{-2}$. This puts a direct lower limit on the quantity $l_{\rm nth}^{\rm max} C^{s_{\rm min}}$. From theoretical arguments, the acceleration slope is constrained to be $s\ge 2$ by Fermi-like acceleration, or at best $s>1$ as suggested by particle-in-cell simulations of magnetic reconnection processes (\citealt{sironi15a, sironi15b,wang15}). Therefore, the maximal compactness parameter must reach at least $l_{\rm nth}^{\rm max}>5\times 10^{-4}$ or $l_{\rm nth}^{\rm max}>1\times 10^{-4}$ respectively, to produce such luminous X-ray flares. This corresponds to a acceleration power of $L_{\rm nth}^{\rm max} > 2.4\times 10^{37}$ and $4.8\times 10^{36}$ erg/s respectively. 

On the other hand, the lower limit of the compactness parameter and the upper limit of the acceleration slope correspond to the lowest fluxes. They suffer from detection limits and flare definition issues and cannot be determined without additional assumptions. 


We conclude that in this scenario, we are able to reproduce quite well the observed IR and X-ray CDFs simultaneously by distributing the non-thermal power and the acceleration slope over the ranges $[5\times 10^{-6} - 1\times 10^{-4}]$ and $[1.1 - 3.6]$ for $l_{\rm nth}$ and $s$, respectively, as $p(l_{\rm nth}) \propto l_{\rm nth}^{-4.2}$ and $p(s)\propto C^{s}$.

\section{SSC scenario}
\label{Results2}

In this section we explore the scenario where the IR emission is produced by non-thermal synchrotron while the X-ray emission partly results from SSC emission by the same population of hot electrons.

It is known that SSC models can reproduce both the quiescent and flaring states of SgrA* for well tuned values of the flow properties \citep[e.g.][]{dibi14}. Compared to the synchrotron-dominated scenario however, the flaring activity implies variations of many physical properties of the accretion flow. There must be changes in not only the properties of particle acceleration, but also the flow density and the average magnetic field. In the following, we assume that the observed variability is due to the fluctuations of 3 parameters only: the acceleration power ($l_{\rm nth}$), the magnetic field ($l_b$), and the plasma density (described by $l_{\rm inj}$). The acceleration slope is assumed to be constant and set to a median value of $s=2.8$. The effect of these three parameters is investigated with a table of spectra computed numerically with the parameters shown in Tab. \ref{tab:SSC_grid1}.
\begin{table}
\begin{center}
\begin{tabular}{c|cccc}
&min &max &number& grid\\ \hline
$l_{\rm nth}$ & $1\times 10^{-6}$ & $5 \times 10^{-3}$ & 100 & log \\
$l_{\rm b}$ & $1\times 10^{-5}$ & $2 \times 10^{-3}$ & 100 & log \\
$l_{\rm inj}$ & $1\times 10^{-4}$ & $2 \times 10^{-2}$ & 50 & log \\
 \hline
\end{tabular}
\caption{Properties of the grids used to produce the table of spectra in the SSC dominated model.} \label{tab:SSC_grid1}
\end{center}
\end{table}

We emphasise here that after investigation of a large parameter space (including other values of the acceleration slope for instance), we were not able to find a realistic model where the X-ray emission is always dominated by Comptonisation, for any value of the varying parameters. It is only the case over limited ranges of parameters and most generally both non-thermal synchrotron and Compton scattering contribute to the X-ray band. 
Moreover, although the NIR band is dominated by non-thermal synchrotron emission, its location relative to the cooling break depends on the exact value of the fluctuating magnetic field. For those reasons, a simple analytical study cannot be performed for this model, contrary to the pure synchrotron scenario, and most constraints must be derived in a numerical approach.

\subsection{The submm band}
\begin{figure}
\includegraphics[scale=0.35]{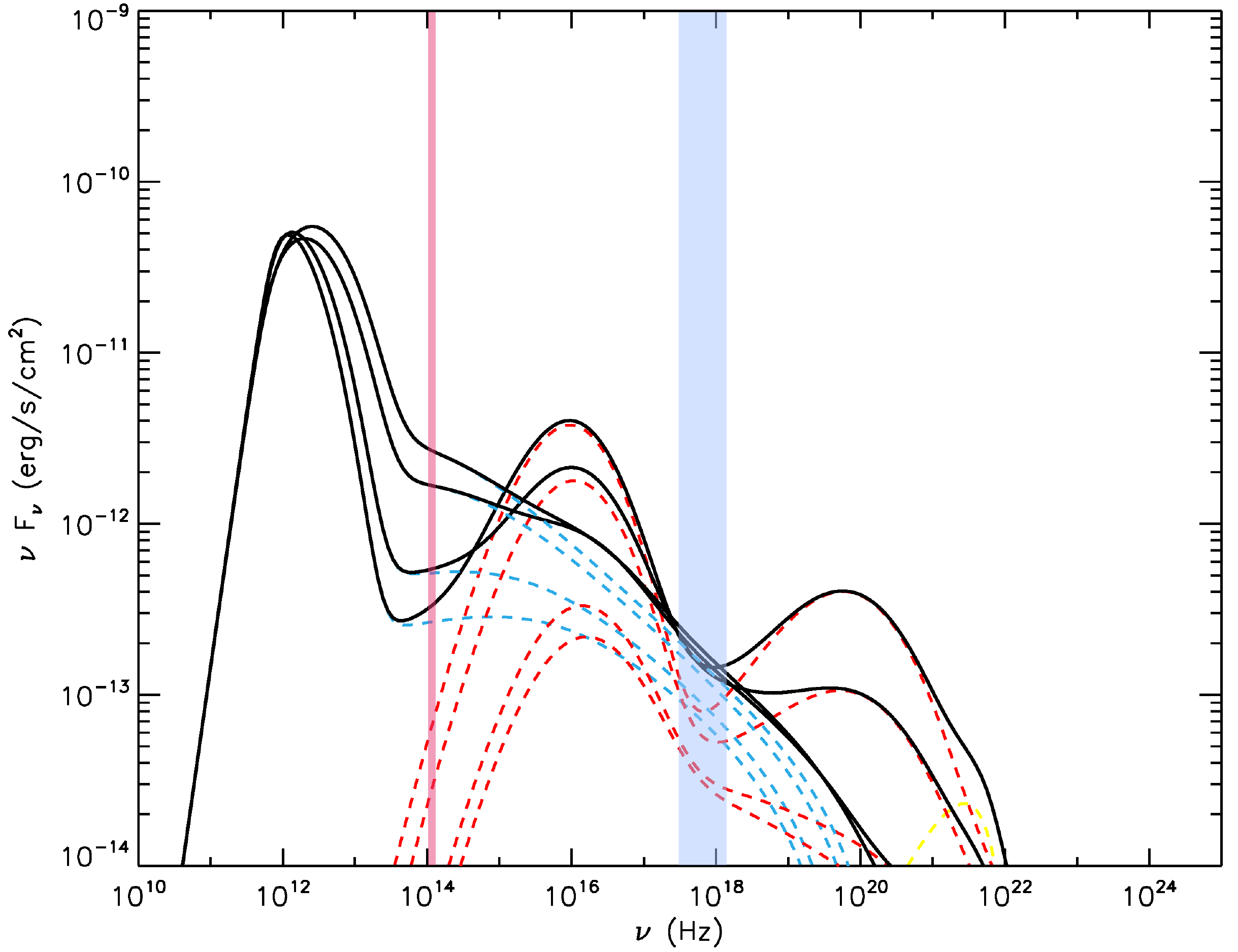}
\caption{\textit{Correlated variations of the magnetic and injection compactness parameters with $l_{\rm inj} = 6\times 10^{-7}/l_b$, $l_{nth}=1.4\times 10^{-5}$, and $s$=2.8.}  Values of $l_{\rm inj}$ correspond to $6.0\times 10^{-4}$, $1.2\times 10^{-3}$, $6.0\times 10^{-3}$, and $1.2\times 10^{-2}$ from top to bottom in the IR band.} 
\label{SPECTRA_B_Correll}
\end{figure}

In contrast to IR and X-rays, the submm bump is observed to be fairly constant, with RMS fluctuations that remain below 20\%. This emission is produced through synchrotron radiation from the thermal part of the lepton population, so that the flux only depends on the flow density $n$ and magnetic field $B$ as: $F_{\rm submm}\propto n B^2$. In terms of the model parameters it is:  $$ F_{\rm submm} \propto l_{\rm inj} l_b  \ .$$ 
In order to keep the submm emission constant, the injection and magnetic compactness parameters must be correlated as $l_{\rm inj} \propto 1/l_b$. This is illustrated in Fig. \ref{SPECTRA_B_Correll} where these two parameters are varied accordingly. The observed fluctuations in the submm band remain consistent with the observational constraints.

\subsection{The X-ray CDF}
Although the Compton and synchrotron contributions to the X-ray band depend on the magnetic field, they happen to produce a steady X-ray flux when $l_b$ and $l_{\rm inj}$ are varied in a correlated manner over a large range of values. This is well illustrated in Fig. \ref{SPECTRA_B_Correll}. The X-ray flux is then simply proportional to the non-thermal population of electrons, i.e., the acceleration power:
$$ F_X \propto l_{\rm nth} \ .$$

\begin{figure}
\centerline{\includegraphics[width=\columnwidth]{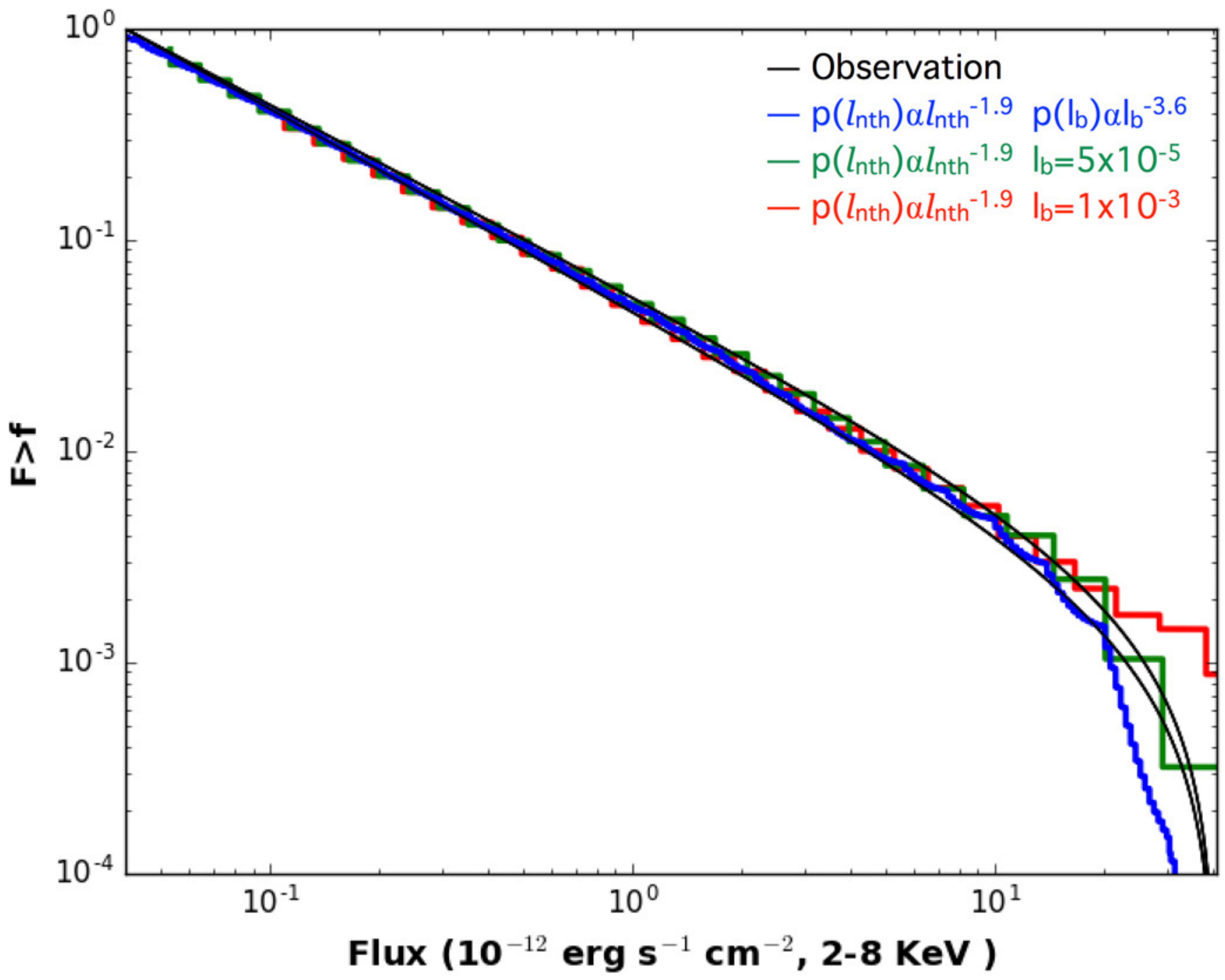}}
\caption{\textit{The blue, green, and red lines are X-ray cumulative flux distributions with random non-thermal acceleration distributed as a power-law $p(l_{nth}) \propto l_{nth}^{-1.9}$. The green and red lines show the CDF for fixed values of the magnetic field ($l_b=5\times 10^{-5}$, and $l_b=1\times 10^{-3}$ respectively). The blue line shows the CDF when the magnetic field is varied independently with distribution $p(l_{b}) \propto l_{b}^{-3.6}$. As in previous plots, the black lines are representative of the observed X-ray CDF.}} 
\label{XCDF_SSC}
\end{figure}

As a consequence, the X-ray CDF is uniquely governed by the fluctuations of the acceleration power. As the relation is linear, the distribution of the non-thermal compactness is a power law with same index as the flux distribution: $p(l_{\rm nth}) \propto l_{\rm nth}^{-1.9}$. The index 1.9 is much smaller than in the synchrotron dominated case. CDFs with this probability distribution are shown in Fig. \ref{XCDF_SSC} for difference values of the magnetic field. As expected, the observed CDF is well recovered, regardless of the magnetic field intensity.

\subsection{The IR CDF}

The IR emission is dominated by non-thermal synchrotron radiation and only depends on non-thermal electron population (i.e. on the non-thermal power) and magnetic field.
More precisely, as long as the magnetic field is weak enough, the cooling break remains at higher energy than the NIR band, and the IR flux is:
$$ F_{IR} \propto l_{\rm nth} l_b^{(s+1)/4} \ . $$
For large fields however, the break appears in the IR band or below. Also, the Comptonisation of the submm bump starts contributing in the IR band for dense flows (i.e. large $l_{\rm inj}$, i.e. weak fields). The exact dependency is therefore more complex. Nevertheless, we find that the IR flux is still well represented by a power-law function of the magnetic field. 

\begin{figure}
\centerline{\includegraphics[width=\columnwidth]{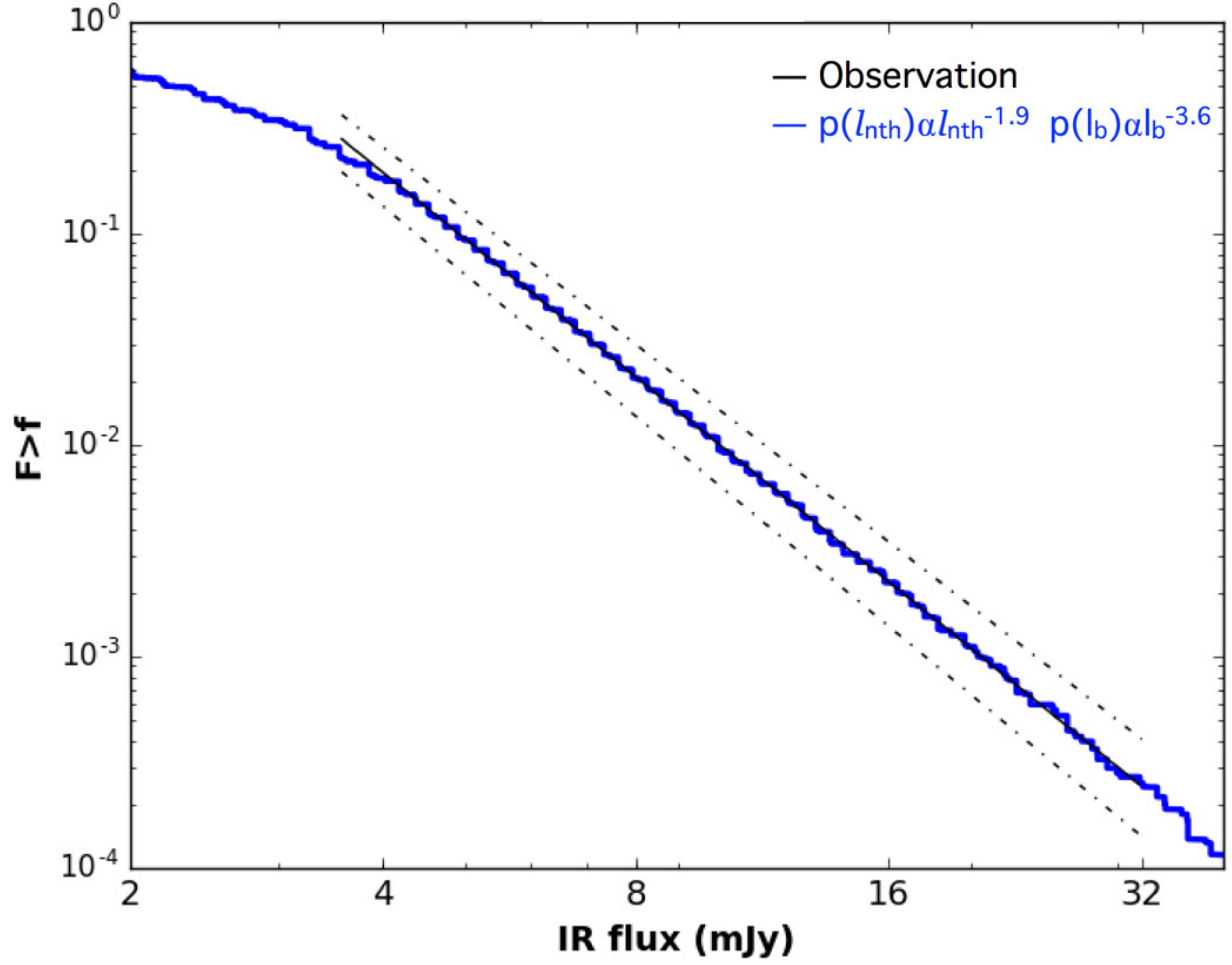}}
\caption{\textit{IR-ray cumulative flux distribution with random non-thermal acceleration distributed as a power-law probability $p(l_{nth}) \propto l_{nth}^{-1.9}$, a random magnetic field distributed as a power-law, $p(l_{b}) \propto l_{b}^{-3.6}$ and an injection compactess parameter correlated to the magnetic field as: $l_{inj}\simeq 6\times 10^{-7}/l_b$.}} 
\label{CDF_SSC_3}
\end{figure}

As the distribution of acceleration power has been derived from the X-ray CDF, the IR CDF can in principle be used to constrain the magnetic field distribution. Such constraint however depends on the correlation between the two parameters. For instance, the magnetic field can be assumed to be independent from the acceleration power. Then we find empirically that, the index $\Gamma$=4.2 of the IR flux distribution can be reproduced with a power-law distribution of the magnetic compactness with index 3.6: $p(l_b) \propto l_b^{-3.6}$. This corresponds to a magnetic field distribution of: $p(B) \propto B^{-6.2}$.
The IR-CDF computed from this probability is shown in Fig. \ref{CDF_SSC_3}.

We see that this scenario can reproduce simultaneously the observed X-ray and IR CDFs slopes fairly well by distributing the non-thermal power and magnetic field over the ranges $[1\times 10^{-6} - 3\times 10^{-4}]$ and $[1.6\times 10^{-5} - 1.6\times 10^{-3}]$ for $l_{\rm nth}$ and $l_b$, respectively, as  $p(l_{\rm nth}) \propto l_{\rm nth}^{-1.9}$ and $p(l_{b}) \propto l_{b}^{-3.6}$ .

\section{Conclusions and discussion}

Our Galactic supermassive black hole Sgr A* exhibits strong IR/X-ray flaring activity, that is often correlated.  This flaring offers a unique opportunity to study the coupling between bulk properties and microphysics. Because Sgr A* is so well studied, we have good constraints on the plasma conditions near the black hole, but the precise emission mechanisms have not been clearly identified. For two families of models: a pure non-thermal synchrotron scenario and a SSC scenario, we have investigated how the flux in the three most commonly observed wavelength bands (submm, IR, and X-ray) responds to fluctuations in the plasma parameters.  

We find that one of the wavelength is always insensitive to one of the parameters expected to govern the flaring activity. This allows us to use the observed CDF in one waveband to infer the distribution of one parameter, and the observed CDF in the other waveband to infer the distribution of the other parameter, with no need to test arbitrary distributions.

In each scenario presented, SD and SSC, the flux dependence on the flow parameters is different in each waveband, which enables us to use the observed flux distributions to derive the statistical behaviour of the flow properties for each case. 

Both the IR and X-ray fluxes are proportional to the power injected in the non-thermal electrons by some acceleration process. Hence the properties of the acceleration process appears to be a key parameter for the flaring activity. We find that the non-thermal power must be distributed as a power-law, index of which depends on the model: in the synchrotron dominated scenario, it is the index of the IR flux distribution, while in the SSC scenario, it is the index of X-ray flux distribution. In the synchrotron dominated model, we found that the non-thermal power must reach at least $4.8\times 10^{36}$ erg s$^{-1}$ to reproduce the brightest flares. The constraints on the other flow parameters depend on their correlation with the non-thermal power. In the synchrotron-dominated model, the flaring activity requires a variation in the acceleration slope. If $s$ is independent of the acceleration power, then it must be distributed as an exponential distribution favouring soft distributions. Although the exact extremal values of the slope could not be derived, we find that $s$ must at least be variable in the range 1.1 to 3.6. Regardless of the acceleration mechanism, it is not clear whether local variations of the flow can produce such a large range of acceleration slopes. 

The submm emission does not show significant variability but depends on the magnetic field and flow density. Hence we find that if each varies with time, their evolution must compensate each other. This implies large variations of the plasma magnetisation. Understanding this correlation is a crucial point to support the SSC scenario, which requires variation of the magnetic field. 

The role of the magnetic field was discussed by \citep{doddseden10}, and while the magnetic field could vary as a result of reconnection or in a turbulent accretion flow, we are not aware of any physical model that would produce this specific correlation.

In summary, we find that both the X-ray and IR flux distributions can be reproduced by pure synchrotron and synchrotron self-Compton models. While we are not able to use observational data to rule out either model, each scenario leaves open questions that may provide avenues for further study. In the synchrotron model, we can reproduce the X-ray and IR flux distributions if s varies significantly over a wide range of slopes, but the broad range required may be difficult to explain with standard acceleration scenarios like diffusive shock acceleration or reconnection. In contrast, the SSC scenario can reproduce the data with a constant acceleration slope, but requires a correlation between the magnetic field and the flow density that has not yet been explained. Thus, more work is required to understand the inferred variations in the acceleration slope and the magnetic field.

\section*{Acknowledgments}

We acknowledge support from The European Community’s Seventh Framework Programme (FP7/2007-2013) under grant agreement number ITN 215212 Black Hole Universe.\\
We also acknowledge support from the ``Nederlandse Onderzoekschool Voor Astronomie'' NOVA Network-3 under NOVA budget number R.2320.0086.\\
This work is partly financed by the Netherlands Organisation for Scientific Research (NWO),  through the VIDI research programme Nr. 639.042.218.\\
S. Markoff is grateful to the University of Texas in Austin for its support, through a Tinsley Centennial Visiting Professorship.\\
R. Belmont and J. Malzac acknowledge financial support from both the french National Research Agency  (CHAOS project ANR-12-BS05-0009) and the Programme National Hautes Energies.\\
J. Neilsen acknowledge support from NASA through Hubble Fellowship grant HST-HF2-51343.001-A.\\
S. Dibi acknowledges fund from an NWO Vidi grant Nr. 2013/15390/EW.

\bibliographystyle{mn2e}
\bibliography{refs}
\bsp

\clearpage

\renewcommand{\thefigure}{\alph{figure}}

\setcounter{figure}{0}

\label{lastpage}
\end{document}